\documentclass[10pt,english,twocolumn,secnumarabic,amssymb,nobibnotes,aps,prl]{revtex4}
\usepackage[T1]{fontenc}
\usepackage[latin9]{inputenc}
\usepackage{amsmath}
\usepackage{graphicx}
\usepackage{amssymb}

\makeatletter

\usepackage{amsfonts}
\usepackage{epsfig}

\usepackage{psfrag}

\usepackage{colordvi}
\usepackage{color}

\newcommand{\mfr}[2]{\left|\begin{smallmatrix} #1\\ #2 \end{smallmatrix}\right\rangle}

\makeatother

\usepackage{babel}

\begin{document}

\title{Two dimensional XXZ-Ising model on square-hexagon lattice}

\author{J. S. Valverde, Onofre Rojas and S. M. de Souza}

\affiliation{Departamento de Ci\^{e}ncias Exatas, Universidade Federal de Lavras.
C.P. 3037, 37200-000, Lavras - MG, Brazil.}

\begin{abstract}
We study a two dimensional XXZ-Ising on square-hexagon (4-6) lattice
with spin-1/2. The phase diagram of the ground state energy is discussed,
shown two different ferrimagnetic states and two type of antiferromagnetic
states, beside of a ferromagnetic state. To solve this model, it could
be mapped into the eight-vertex model with union jack interaction
term. Imposing exact solution condition we find the region where the
XXZ-Ising model on 4-6 lattice have exact solutions with one free
parameter, for symmetric eight-vertex model condition. In this sense
we explore the properties of the system and analyze the competition
of the interaction parameters providing the region where it has an
exact solution. However the present model does not satisfy the \textit{free
fermion} condition, unless for a trivial situation. Even so we are
able to discuss their critical points region, when the exactly solvable
condition is ignored. 
\end{abstract}
\maketitle
\sloppy 

Recently frustrated magnetic systems have been attracting a lot of
attention due to their rich properties. Such systems have several
phase diagrams displaying a number of unusual quantum phases\cite{ritcher-prl,ritcher-lnp}.
Frustration interaction is exhibit experimentally in inelastic neutron
scattering. Then two dimensional magnetic lattice is a challenge for
the theoretical investigation. After Onsager's\cite{onsager} solution
of the square two dimensional Ising lattice, other solutions for regular
two-dimensional lattices, such as triangular\cite{newell,husimi},
honeycomb\cite{husimi1}, Kagom\'{e}\cite{syozi} lattice and others
were explored in several works and his importance in statistical physics
waked up to search for a group of completely solvable models. The
problem concerning to the exact solution and the critical behavior
of the two-dimensional models was the matter of the Fan and Wu\cite{fan1}-\cite{wu}.
In those works the \textit{free fermion} (FF) condition and the free
fermion approximation was studied with great details and the relations
of the Boltzmann weights for obtaining exact solvable models was established.
In many situations when the FF condition is not satisfied completely
it is possible to find with a good approximation for particular values
of the parameters of the model. This is the case investigated by Kun-Fa\cite{kun}
where the critical coupling of mixed Ising spin-1/2 with the arbitrary
Ising spin-S was studied using the free fermion approximation. 

Since that many theoretical investigation was developed, such as Ising-Heisenberg
Kagom\'{e} lattice\cite{ma1,jascur08}, quantum square-Kagom\'{e} antiferromagnetic
lattice\cite{siddharthan}, doubly decorated Ising-Heisenberg model\cite{jascur02},
the mixed-spin Ising model on a decorated square lattice with two
different kinds of decorating spins on horizontal and vertical bonds\cite{jascur07}.
Other exactly solvable Ising model lattice known as square-hexagon
(4-6) was considered by Lin an Yang\cite{lin-yang}. On the other
hand a different 4-6 lattice as a special case of the 4-8 lattice
was studied by Oitmaa and Keppert\cite{keppert} where the solution
for the Ising spin-1/2 case was found. It is remarkable to point out
that the free fermion condition for the Boltzmann weights is satisfied
identically in these models and the exact critical point can be performed,
thus the models falls within the standard Ising universality class.

Several real systems motivate to investigate in this kinds of lattice,
such as the recently discovered two-dimensional magnetic materials
$Cu_{9}X_{2}(cpa)_{6}.xH_{2}O$ (cpa=2-carboxypentonic acid; X=F,
Cl, Br) where the $Cu$ spins stands on the triangular Kagom\'{e} lattice\cite{norman}
with Heisenberg interaction type. Liquid crystals networks composed
by pentagonal, square and triangular cylinders\cite{bin-chen}. Other
recent investigation about the crystal structure of solvated {[}Zn(tpt)2/3(SiF,)(H20)2-
(MeOH)] {[}tpt = 2,4,6-tris(4-pyridyl)-1,3,5-triazine] networks with
the (10,3)-a topology\cite{robson}.

\begin{widetext}

\begin{figure}[!ht]
\begin{centering}
\psfrag{a}{(a)}\psfrag{b}{(b)}\psfrag{c}{(c)} \psfrag{s1}{$\sigma_{1}$}
\psfrag{s2}{$\sigma_{2}$} \psfrag{s3}{$\sigma_{3}$}\psfrag{s4}{$\sigma_{4}$} \psfrag{K}{$K$}
\psfrag{L}{$L$} \psfrag{M}{$M$} \psfrag{Hi}{$J$} \psfrag{Hxxz}{$(J,,J_{z})$}
\psfrag{s}{$s$} \psfrag{m}{$\mu$}\psfrag{S1}{$S_{1}$} \psfrag{S2}{$S_{2}$}
\psfrag{S3}{$S_{3}$}\psfrag{S4}{$S_{4}$} \includegraphics[width=14cm,height=4cm]{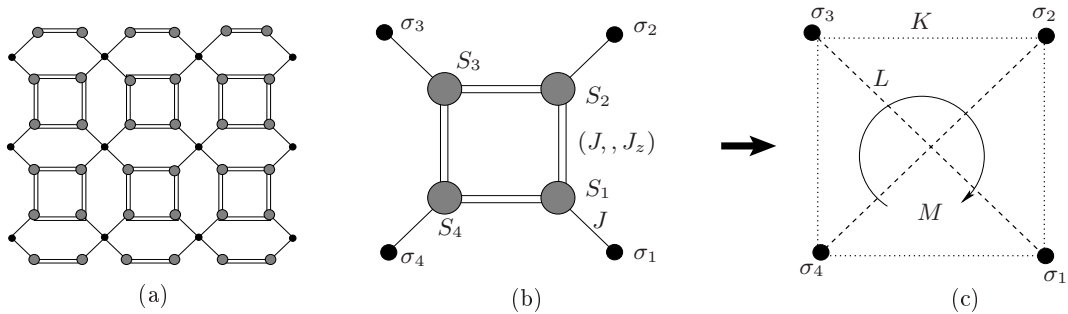} 
\par\end{centering}

\caption[fig5]{Schematic representation of two dimensional XXZ-Ising on 4-6 lattice.
In (a) we represent as double line the XXZ interaction whereas with
single line we represent the Ising interaction. In (b) we illustrate
each decorated cell displayed in (a). The transformation of unitary
cell is represented in (c), only as a function of the Ising interaction
with nearest, next-nearest and quartic interaction of effective Ising
lattice.}

\label{fig_1} 
\end{figure}

In this report we will discuss a similar model presented in ref. \cite{keppert},
where we include local Heisenberg interaction. In Fig.\ref{fig_1}a
we display schematically two dimensional XXZ-Ising on 4-6 lattice,
with a single line we represent the Ising lattice, whereas with double
line we indicate the XXZ interaction. In fig. \ref{fig_1}b we represent
each decorated cell of the lattice, therefore the Hamiltonian of XXZ-Ising
lattice will become as\begin{align}
{\mathcal{H}}(\{\sigma\},\{S^{\alpha}\})=\sum_{i}JS_{i}\sigma_{i}+\sum_{<i,j>}\Big(J(S_{i}^{x}S_{j}^{x}+S_{i}^{y}S_{j}^{y})+J_{z}S_{i}^{z}S_{j}^{z}\Big),\label{decorado}\end{align}
 \end{widetext} the first summation runs over all sites involving
the Pauli operator $\sigma$ with two possible values $\pm1$, while
the second summation runs over nearest neighbor of whole lattice containing
the $S^{\alpha}$ spin-1/2 operators with $\alpha=\{x,y,z\}$.

\begin{figure}[!ht]
\psfrag{Jz}{$J_{z}$} \psfrag{J}{$J$} \psfrag{FM}[][][0.5]{$|FM\rangle$}
\psfrag{AF1}[][][0.5]{$|AF_1\rangle$} \psfrag{AF2}[][][0.5]{$|AF_2\rangle$}
\psfrag{FI1}[][][0.5]{$|FI^{-}\rangle$} \psfrag{FI2}[][][0.5]{$|FI^{+}\rangle$}

\begin{centering}
\includegraphics[width=5cm,height=7cm,angle=-90]{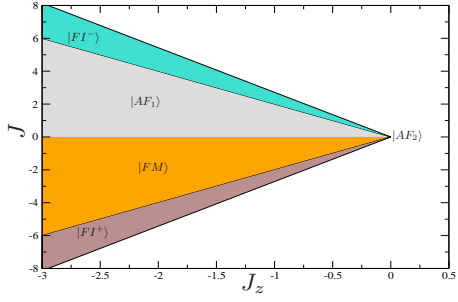} 
\par\end{centering}

\caption[fig1]{The phase diagram at zero temperature, as a function of two parameters,
$J_{z}$and $J$.}

\label{fig_3} 
\end{figure}

The phase diagram at zero temperature for the two dimensional XXZ-Ising
on 4-6 lattice is analyzed, computing the ground state energy. We
verify that there are five different states, as depicted in fig.\ref{fig_3}.
We can classify the states by ferromagnetic and antiferromagnetic
in the Ising interaction part. Therefore let us start writing the
ground state energy when the Ising interaction is ferromagnetic,\begin{align}
|FM\rangle & =\mfr{^{+}++^{+}}{_{+}++_{+}},\label{FM}\\
|AF_{1}\rangle & =\mfr{^{-}++^{-}}{_{-}++_{-}},\label{AF1}\\
|FI^{\pm}\rangle & =\sum_{r=0}^{3}\left(\pm\mathsf{R}\right)^{r}\mfr{^{\pm}++^{\pm}}{_{\pm}-+_{\pm}},\label{FIM}\end{align}
by the large $\pm$ (inner squared signals) we represent the Heisenberg
interaction particles, whereas by the corner small signals $\pm$
we indicate the Ising interaction particles. By $\mathsf{R}$ we represent
the rotation operator acting only in Heisenberg interaction particles,
each rotation is performed in $\tfrac{\pi}{2}$, around of the axis
perpendicular to the plane of lattice.

The ferromagnetic state ($|FM\rangle$) given in eq.\eqref{FM} for
$J_{z}<0$, is limited by $\tfrac{2}{5}J_{z}<J<0$, this state is
depicted as brown region in fig.\ref{fig_3}. There is also an antiferromagnetic
state ($|AF_{1}\rangle$) represented by eq. \eqref{AF1} for $J_{z}<0$,
which is restricted in the interval $0<J<-\tfrac{2}{5}J_{z}$, and
is displayed in fig.\ref{fig_3} as gray region. It is worth to notice
that the state behaves as ferromagnetic for both in Ising and Heisenberg
interaction particles but with opposite oriented spin. Under same
condition we also have two types of ferrimagnetic states (FI) which
are explicitly given by eq.\eqref{FIM}, for $J_{z}<0$, the $|FI^{+}\rangle$
is the ferrimagnetic state with magnetization $3/4$, limited by $\tfrac{2}{5}J_{z}<J\lesssim0.43067503J_{z}$,
whereas the $|FI^{-}\rangle$ corresponds to the ferrimagnetic state
with magnetization $1/4$ restricted by $-0.43067503J_{z}\lesssim J<-\tfrac{2}{5}J_{z}$.
Those regions are illustrated in fig.\ref{fig_3} as orange and cyan
region respectively. These ferrimagnetic states are invariant under
whole exchange of spin orientation.

The other possible situation is when Ising interaction antiferromagnetic
is considered. In this situation we only have one antiferromagnetic
state $|AF_{2}\rangle$, we also could call this state as frustrated
state\cite{v-jpcm08}. This ground energy is obtained after diagonalized
a $3\times3$ matrix, which is fall into a cubic equation and their
lowest solution is given by $-\tfrac{8}{3}\left(P_{1}\cos(\phi_{1})-J_{z}\right)$,
with

\begin{align}
P_{1}= & \sqrt{J_{z}^{2}+9J^{2}},\label{def-P1}\\
\phi_{j}=\tfrac{1}{3} & \cos^{-1}\left(\tfrac{J_{z}^{3}}{P_{1}^{3}}\right)+\tfrac{2\pi j}{3},\label{cubic-sol1}\end{align}
in eq.\eqref{cubic-sol1}, $\phi_{j}$ (with $j=0,1,2$) is related
to the real root of the cubic equation. The corresponding eigenvector
state read as \begin{align}
|AF_{2}\rangle & =b_{1}\sum_{r=0}^{3}\mathsf{R}^{r}\mfr{^{+}++^{-}}{_{-}--_{+}}+(1+b_{2}\mathsf{R)}\mfr{^{+}+-^{-}}{_{-}-+_{+}},\label{AF2}\end{align}
and the coefficients of \eqref{AF2} are given by\begin{align}
b_{1}= & \tfrac{1}{6J}\left(2P_{1}\cos(\phi_{1})+J_{z}+3J\right),\\
b_{2}= & \tfrac{4}{3J}b_{1}\left(P_{1}J\cos(\phi_{1})-J_{z}\right)-1.\end{align}

This antiferromagnetic state is present for arbitrary values of $J$
when $J_{z}>0$, whereas for $J_{z}<0$ this states is limited by
$|J|\gtrsim-0.43067503J_{z}$, as illustrated in fig. \ref{fig_3}.
We remark that in this case we have antiferromagnetic interaction
for both Ising and XXZ interactions.

It is worth to comment that this model has multicritical point at
zero temperature, where five states converge at $J=J_{z}=0$. 

To study the thermodynamics of the model we write the partition function
of decorated XXZ-Ising lattice, given by Hamiltonian \eqref{decorado},\begin{align}
\mathcal{Z}(\beta)=\sum_{\{\sigma\}=\pm1}{\rm tr}_{\{S^{\alpha}\}}\Big({\rm e}^{-\beta{\mathcal{H}}(\{\sigma\},\{S^{\alpha}\})}\Big).\end{align}

After taking the trace over operators $\{S^{\alpha}\}$ we transform
the decorated XXZ-Ising model into an effective Ising model, with
next nearest and quartic interactions parameter, whose effective Hamiltonian
could be expressed in general by\begin{align}
\widetilde{\mathcal{H}}(\{\sigma\})=K\sum_{<i,j>}\sigma_{i}\sigma_{j}+L\sum_{(i,j)}\sigma_{i}\sigma_{j}+M\sum_{\substack{\text{all}\\
\text{square}}
}\sigma_{1}\sigma_{2}\sigma_{3}\sigma_{4},\label{efetivo}\end{align}
 with $K$ being the nearest neighbor interaction, whereas $L$ is
the next nearest neighbor interaction parameter and $M$ being the
quartic interaction parameter. This transformation is also represented
schematically in fig.\ref{fig_1}c. This effective Ising model is
the so called 'Union Jack' lattice, which is an exactly solvable model\cite{baxter-c}.

Therefore the corresponding partition function of effective Ising
lattice is given by\begin{align}
\mathcal{Z}_{eff}=f\sum_{\{\sigma\}=\pm1}\Big({\rm e}^{-\beta\widetilde{\mathcal{H}}(\{\sigma\})}\Big).\end{align}

\begin{figure}[h]
\begin{centering}
\psfrag{m}{$-$} \psfrag{p}{$+$} \psfrag{w1}{$w_{1}$} \psfrag{w2}{$w_{2}$}
\psfrag{w3}{$w_{3}$} \psfrag{w4}{$w_{4}$} \psfrag{w5}{$w_{5}$}
\psfrag{w6}{$w_{6}$} \psfrag{w7}{$w_{7}$} \psfrag{w8}{$w_{8}$}
\includegraphics[width=8cm,height=1cm]{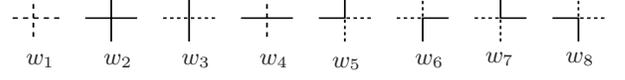} \caption[fig1]{The eight spin vertex configurations. Reversal of all spins corresponds
to the same vertex}
\label{fig_2} 
\par\end{centering}
\end{figure}

Using the Boltzmann weight given in fig.\ref{fig_2} we are able to
transform the Hamiltonian \eqref{decorado} into \eqref{efetivo}
where their parameters are related by the following expression,\begin{align}
f= & \big(w_{1}w_{2}w_{3}^{2}w_{5}^{4}\big)^{\frac{1}{8}},\\
K= & -\frac{1}{8\beta}\ln\Big(\frac{w_{1}}{w_{2}}\Big),\\
L= & -\frac{1}{8\beta}\ln\Big(\frac{w_{1}w_{2}}{w_{3}^{2}}\Big),\\
M= & -\frac{1}{8\beta}\ln\Big(\frac{w_{1}w_{2}w_{3}^{2}}{w_{5}^{4}}\Big).\end{align}

Performing some algebraic manipulation we write the associated Boltzmann
weights of the Hamiltonian \ref{decorado}. This large results are
written using some extra notations just to express in a compact form,\begin{widetext}
\begin{align}
w_{1}= & 3+\mathrm{e}^{4\beta J_{z}}+8\cosh^{3}(2\beta J)+2{\rm \mathrm{e}^{-4\beta J_{z}}}\cosh(4\beta J)+2{\rm \mathrm{e}^{-2\beta J_{z}}}\cosh\left(2\beta\sqrt{J_{z}^{2}+8J^{2}}\right),\label{bo1}\\
w_{2}= & 3+2{\rm e}^{-4\beta J_{z}}+4\cosh(2\beta J)+4\cosh(2\sqrt{5}\beta J)+\sum_{j=0}^{2}\mathrm{e}^{-\frac{8}{3}\beta\left(P_{1}\cos\left(\phi_{j}\right)-J_{z}\right)},\label{bo2}\\
w_{3}= & 1+\mathrm{e}^{4\beta J_{z}}+2{\rm \mathrm{e}^{-4\beta J_{z}}}+8\cosh(2\sqrt{2}\beta J)\cosh(2\beta J)+2{\rm e}^{\frac{\beta}{4}J_{z}}({\rm e}^{\beta P_{2}}\cosh(\beta A_{+})+{\rm e}^{-\beta P_{2}}\cosh(\beta A_{-})),\label{bo3}\\
w_{5}= & 2+\sum_{j=0}^{2}\cosh\left(\tfrac{4\beta J}{3}\left(4\cos\left(\theta_{j}\right)-1\right)\right)+2{\rm e}^{2\beta J_{z}}\big(\cosh(\beta B_{+})+\cosh(\beta B_{-})\big)+2\cosh(2\beta J)(1+{\rm e}^{-4\beta J_{z}}),\label{bo4}\end{align}

\end{widetext} with $P_{1}$ and $\phi_{j}$ already was defined
in eqs.\eqref{def-P1} and \eqref{cubic-sol1} respectively, whereas
the other amounts are defined as follow \begin{align}
\theta_{j}= & \tfrac{1}{3}\cos^{-1}\left(\tfrac{5}{32}\right)+\tfrac{2\pi j}{3},\\
P_{2}= & \sqrt{J_{z}^{2}+10J^{2}+2J\sqrt{4J_{z}^{2}+25J^{2}}},\\
A_{\pm}= & \sqrt{(3\pm\tfrac{2J_{z}}{P_{2}})(J_{z}^{2}+4J^{2})+12J^{2}-P_{2}^{2}},\\
B_{\pm}= & 2\sqrt{J_{z}^{2}+5J^{2}\pm2J\sqrt{6J^{2}+J_{z}^{2}}}.\end{align}
The other Boltzmann weights can be obtain using the symmetry rotation,
thus we have the following identities for the model considered,\begin{align}
w_{3}=w_{4},\quad\text{and}\quad w_{5}=w_{6}=w_{7}=w_{8}.\end{align}

The two dimensional XXZ-Ising on 4-6 lattice model in general has
no exact solution, but it could be possible to find some particular
solutions imposing the exact solvable condition. Therefore it is possible
to find one branch region where the model have an exact solution.

The first branch of a possible exact solution could be when the Boltzmann
weights satisfy the so called symmetric eight-vertex model (SEVM)
condition, where we must have the following relations,\begin{align}
w_{1}=w_{2},\quad w_{3}=w_{4},\quad w_{5}=w_{6}\quad\text{and}\quad w_{7}=w_{8}.\label{sevmc}\end{align}

Our model satisfy all these relations given by \eqref{sevmc}, unless
the first one.

\begin{figure}[!ht]
\psfrag{bJ}[][][0.7]{$\beta J$} \psfrag{bJz}[][][0.7]{$\beta J_{z}$}
\psfrag{Dlt>10^-3}[][][0.7]{$\Delta>10^{-3}w_{max}^2$}\psfrag{Dlt<10^-3}[][][0.7]{$\Delta<10^{-3}w_{max}^2$}

\begin{centering}
\includegraphics[width=5cm,height=8cm,angle=-90]{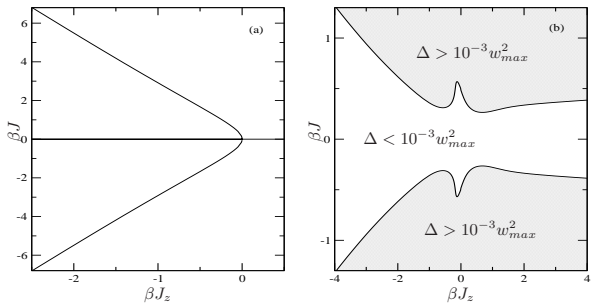} 
\par\end{centering}

\caption{In (a) is displayed the exactly solvable condition for SEVM ($w_{1}=w_{2}$).
(b) The FF is imposed and there is no exact solution. But there is
a valley where $\Delta/w_{max}^{2}\ll1$, particularly we show the
region for $\Delta/w_{max}^{2}>10^{-3}$ as gray region.}

\label{fig_4} 
\end{figure}

Imposing the first relation of eq. \eqref{sevmc} we have one possible
solution. In fig.\ref{fig_4}(a) we display the exactly solvable SEVM
condition, as function of the parameters $J$ and $J_{z}$ in units
of $\beta$. Therefore we show that for one free parameter the Hamiltonian
\eqref{decorado} could be solved exactly, spite this transcendental
equation evolves complicated relation of $J_{z}$ and $J$, we are
not able to invert one of them as a function of the other one explicitly,
but even so we can invert numerically. In the limit for large values
of $J_{z}$ and $J$, we have the asymptotic limit where the relation
becomes approximately by $J\thickapprox\pm2.713579J_{z}$. We also
have a trivial solution when $J=0$, this corresponds just to a set
of non-interacting squared Ising model. 

The second candidate for the exact solution is the so called \textit{free
fermion} (FF) condition, when the following relation\begin{equation}
\Delta=w_{1}w_{2}+w_{3}w_{4}-w_{5}w_{6}-w_{7}w_{8},\label{in1}\end{equation}
 must satisfy the condition $\Delta=0$.

Imposing the FF condition unfortunately we cannot find a solution
for this model, unless for a trivial condition when $J=0$. In this
situation we always have a positive amount of $\Delta/w_{max}^{2}$.
Then we can note that, if we display for small $\Delta/w_{max}^{2}$
the relation of $J_{z}$ and $J$ in units of $\beta$, there are
a wide valley where satisfy the condition $\Delta/w_{max}^{2}\ll1$,
particularly we display in fig.\eqref{fig_4}, a valley for $\Delta/w_{max}^{2}<10^{-3}$,
whereas the gray region correspond to the condition of $\Delta/w_{max}^{2}>10^{-3}$.
This means that we can approximate to the FF condition and solve this
model with good approximation in all this region. 

\begin{figure}[!ht]
\psfrag{bJ}{$\beta J$}\psfrag{bJz}{$\beta J_{z}$}\psfrag{w1>w2 (FFC)}[][][0.5]{$w_1>w_2$ (FF)}

\psfrag{w1<w2 (FFC)}[][][0.5]{$w_1<w_2$ (FF)} \psfrag{w1=w2 (SEVMC)}[][][0.5]{$w_1=w_2$ (SEVM)}

\begin{centering}
\includegraphics[width=5cm,height=7cm,angle=-90]{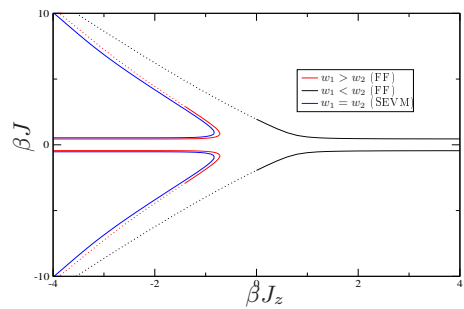} 
\par\end{centering}

\caption[fig1]{The critical points region under FF condition: red line corresponds
the condition of $w1>w2$, whereas by black line we represent the
condition when $w1<w2$. Doted line corresponds the region where $\Delta/w_{max}^{2}>1$.
On the other hand when SEVM condition is imposed, the critical region
becomes the curve given by blue line.}

\label{fig_5} 
\end{figure}

It is also possible to discuss the critical behavior, even when exactly
solvable condition is not satisfied. For the first branch solution
(SEVM), its critical condition must satisfy the following relation\begin{equation}
w_{1}+w_{3}+w_{5}+w_{7}=2\text{max}(w_{1},w_{3},w_{5},w_{7}).\label{in-2a}\end{equation}
In fig.\ref{fig_5} we display the critical points region as a function
of the parameters $J$ and $J_{z}$ in units of $\beta$, and we represent
by a solid blue line. The convergence for this case is satisfied in
all critical points $|\Delta'|/w_{max}^{2}<1$, with $|\Delta'|=|w_{1}-w_{2}|$
and $w_{max}=max\{w_{1},w_{2}\}$. 

The second branch critical points region is when we impose the FF
condition\begin{equation}
w_{1}+w_{2}+w_{3}+w_{4}=2\text{max}(w_{1},w_{2},w_{3},w_{4}).\label{in-1a}\end{equation}
In fig.\ref{fig_5} we display the critical points region as a function
of the parameters $J$ and $J_{z}$ in units of $\beta$, the case
when the Boltzmann weight $w_{1}$ is taken as the maximum value.
The red solid lines indicates the region where the FF approximation
is valid ($|\Delta|/w_{1}^{2}<1$), and the doted red line indicates
the region where $|\Delta|/w_{1}^{2}>1$. The black solid line display
the critical condition region when the $w_{2}$ is the largest one,
with restriction $\Delta/w_{2}^{2}<1$, while with doted black line
we represent the critical region when $\Delta/w_{2}^{2}>1$.

In this report we discuss some particular solution of two-dimensional
XXZ-Ising model on square-hexagon lattice, where the decoration is
a square with XXZ interaction and the interaction terms of the lattice
is given by Ising type coupling. We discuss the phase diagram at zero
temperature displaying five different phases. To study their thermodynamic,
initially we used two parameters but due to imposing the exact solvable
condition we constrain this two parameters, then we obtain a 2D XXZ-Ising
on 4-6 lattice with one free parameter, under SEVM condition. Under
FF condition, we display a wide valley where the model could be considered
approximately as satisfying the FF condition. It is also possible
to discuss the critical condition even when the exact result condition
is not satisfied.

J. S. V. Thanks FAPEMIG for full financial support. O. Rojas. and
S.M. de Souza. thanks CNPq and FAPEMIG for partial financial support.

\end{document}